\def\year{2022}\relax
\documentclass[letterpaper]{article} 
\usepackage{aaai22}  
\usepackage{times}  
\usepackage{helvet}  
\usepackage{courier}  
\usepackage[hyphens]{url}  
\usepackage{graphicx} 
\usepackage{natbib}  
\usepackage{caption} 
\DeclareCaptionStyle{ruled}{labelfont=normalfont,labelsep=colon,strut=off} 
\usepackage{algorithm}
\usepackage{algorithmic}

\usepackage[utf8]{inputenc} 
\usepackage[T1]{fontenc}    
\usepackage{hyperref}       
\usepackage{url}            
\usepackage{booktabs}       
\usepackage{amsfonts}       
\usepackage{nicefrac}       
\usepackage{microtype}      
\usepackage{xcolor}         
\usepackage{mathtools}
\usepackage{subfig}
\usepackage{newfloat}
\usepackage{listings}
\floatstyle{ruled}
\newfloat{listing}{tb}{lst}{}
\floatname{listing}{Listing}
\title{Interpretable Modeling of Single-cell perturbation Responses to Novel Drugs Using Cycle Consistence Learning}
\author{
   Wei Huang, Aichun Zhu, Hui Liu*
}
\affiliations{

    \textsuperscript{\rm 1} College of Computer and Information Engineering, Nanjing Tech University, Nanjing, 211816, China.\\

    *Corresponding author. E-mail(s): hliu@njtech.edu.cn
}

\usepackage{bibentry}

\begin{document}

\maketitle

\begin{abstract}
 Phenotype-based screening has attracted much attention for identifying cell-active compounds. Transcriptional and proteomic profiles of cell population or single cells are informative phenotypic measures of cellular responses to perturbations. In this paper, we proposed a deep learning framework based on encoder-decoder architecture that maps the initial cellular states to a latent space, in which we assume the effects of drug perturbation on cellular states follow linear additivity. Next, we introduced the cycle consistency constraints to enforce that initial cellular state subjected to drug perturbations would produce the perturbed cellular responses, and, conversely, removal of drug perturbation from the perturbed cellular states would restore the initial cellular states. The cycle consistency constraints and linear modeling in latent space enable to learn interpretable and transferable drug perturbation representations, so that our model can predict cellular response to unseen drugs. We validated our model on three different types of datasets, including bulk transcriptional responses, bulk proteomic responses, and single-cell transcriptional responses to drug perturbations. The experimental results show that our model achieves better performance than existing state-of-the-art methods. 
\end{abstract}

\section{Introduction}
Notwithstanding target-based drug discovery has made substantial advancements in recent years, intervention of a specific target (protein or RNA) by compounds has been insufficient in establishing the systematic correlation with organism-level therapeutic effects or side effects. Consequently, the failure rate of leading compounds generated from target-based screening to approved drugs remains high. As such, there has been a renewed interest in phenotypic drug discovery for the identification of cell-active compounds.

Transcriptional profiles serve as a robust and informative phenotypic measure of cellular responses to perturbations. Large-scale compendia have been established to examine the drug-induced phenotypic alterations across cancer cell lines, including large-scale pharmacologic perturbation studies and cell viability measurements upon different drug treatments. For example, the L1000 platform \cite{subramanian2017next} has been developed for high-throughput profiling of mRNA responses of cancer cell lines to diverse perturbations. In parallel, proteomic responses of cancer cell lines to a diverse array of clinically relevant drugs have also been generated using the reverse-phase protein arrays (RPPAs). The cancer perturbed proteomics atlas (CPPA) profiled large-scale drug responses of more than 200 clinically relevant proteins that covered major targets for cancer therapy \cite{zhao2020large}. Transcriptional and proteomic profiling reflect the multi-level regulatory state transition upon external perturbations, providing sound measurements of cellular response that greatly facilitate phenotype-based drug screening.

The single-cell RNA sequencing (scRNA-seq) can identify subtle changes in gene expression and tumor heterogeneity at single-cell resolution, which is important for distinguishing distinctive effects of certain perturbation on cell subpopulations and identifying cellular subsets resistant to specific drug. For example, sci-Plex \cite{srivatsan2020massively} uses nuclear hashing to quantify global transcriptional responses to thousands of independent perturbations at single-cell resolution. The single-cell profiling enables interrogation of phenotypic heterogeneity at a level which has been hitherto inaccessible. However, these high-throughput screening technologies is still limited relative to the vast combinatorial landscape of all cell type (cell number)-perturbation pairs. Therefore, computational model trained on the observed experimental data to predict cellular responses within various cellular contexts is of great importance.

In particular, the model capacity to predict cellular responses to unseen perturbations \cite{yu2022perturbnet}, particularly useful for drug repurposing, may have significant medical implications. This requires the model to effectively capture the intricate interactions between chemical components and cellular molecules that may intrigue cascade biochemical reactions and eventually drive the transition of molecular phenotype. The cellular response to drug perturbation is controlled by underlying biological network, and the cellular state transition is actually nonlinear. To build an interpretable model, we assume the effects of drug perturbation on cellular states follow linearly additive rule. Based on this assumption, we introduced the cycle consistency constraints to enforce that initial cellular state subjected to drug perturbations would produce the expected cellular responses, and, conversely, removal of drug perturbation from the perturbed cellular states would restore the initial cellular states. The cycle consistency constraints and linear modeling enable to learn interpretable and transferable drug perturbation representations, so that our model can predict cellular response to unseen drugs. We validated our model on three different types of datasets, including bulk transcriptional drug responses, bulk proteomic drug responses and single-cell drug responses. The experimental results show that our model achieves better performance than existing state-of-the-art methods.

We think this work has at least three contributions as below:
\begin{itemize}
    \item To our best knowledge, we are the first to introduce cycle consistency loss into learning cellular responses to drug perturbations, which enables our model to learn expressive and transferable drug representations.
    \item We model the cellular response from two opposite perspectives, requiring the model to simultaneously predict the cell state transitions from unperturbed state (control) to perturbed state (treatment) and vice versa, which enforce the encoder networks to capture the essential feature of drug perturbations to cellular state.
    \item We not only evaluate the proposed model on both bulk and single-cell transcriptional responses, but also introduce a proteomic drug response dataset to evaluate our model. The experimental results on these benchmark datasets demonstrate the superior performance of our model. To our best knowledge, this is the first to apply proteomic data to evaluate prediction model of cellular response to novel drug perturbation.
\end{itemize}

\section{Related Works}

\subsection{Predicting cellular response to perturbation}
Some computational methods have been developed to predict cellular responses to perturbations. Among them, mechanistic modeling has been leveraged to predict cell viability or the abundance of specific proteins. While these models are powerful at interpreting interactions, they typically require longitudinal data, which is often unavailable in practice. Furthermore, most mechanistic models do not scale well to genome-wide measurements or high-dimensional scRNA-seq data, making them less suitable for predicting high-dimensional responses. However, thanks to the development of deep learning technology in recent years, this tool has been increasingly applied in the analysis and interpretation of scRNA-seq data \cite{hetzel2021graph, lopez2020enhancing}.

Some machine learning methods that have been proposed for predicting cellular responses to drug treatments. These methods include deep variational autoencoder \cite{jia2021deep}, kernelized Bayesian matrix factorization \cite{madhukar2019bayesian}, matrix factorization with similarity regularization \cite{gao2021collaborative}, convolutional neural network \cite{zhang2022cnn}. These methods leverage different techniques such as imputing drug response through low embedding of multiple genes \cite{roohani2022gears}, incorporating prior knowledge of pathway-drug associations \cite{chawla2022gene}, leveraging mutational signatures \cite{robichaux2021structure, aissa2021single}, and using gene expression data for prediction \cite{sharifi2021out, he2022context}.

\subsection{Linear model in latent space}
The linear additive model in the latent space is widely used in deep learning for interpretability. Specifically, these models use matrix factorization techniques or deep generative models to predict drug response using linear models in a low-dimensional latent space representation. CPA \cite{lotfollahi2021learning} and chemCPA \cite{hetzel2022predicting} are most related to our work, as they combine the interpretability of linear models with the flexibility of deep-learning approaches for single-cell response modeling. Although these models generate easy-to-interpret embeddings for drugs and cells, their accuracy is still insufficient to drive drug discovery.

\subsection{Cycle consistency}
Cycle consistency is a concept used in many computer vision problems that involve processing multiple entities. It is a way of seeking global agreement by enforcing consistency between local relationships. One common way of enforcing cycle consistency is through the use of a cycle consistency loss, which is a type of loss function that encourages forward and backward consistency between mappings. It was first proposed in CycleGAN \cite{zhu2017unpaired}. In this essay, Cycle consistency refers to the property that the image output by the first generator can be used as input to the second generator, and the output of the second generator should match the original image. The reverse is also true. This reduces the space of possible mapping functions and helps to ensure that the mappings are consistent with each other. We are the first to apply cycle consistency in predicting cellular responses to drug perturbation, which enable our model to learn informative and transferable drug perturbation features within various cellular contexts.

\subsection{Domain adaptation}
Domain adaptation refers to the process of adapting a model from the source domain to the target domain, where the data distributions of the source and target domains are different. For general domain adaptation methods, they can be classified into methods based on domain distribution differences \cite{tzeng2019deep,chen2019joint,arjovsky2017wasserstein}, adversarial learning-based methods \cite{wang2020continuously,pei2018multi,long2018conditional}, reconstruction-based methods \cite{zheng2017learning}, and sample generation-based methods \cite{sankaranarayanan2018generate}. In predicting drug response, domain adaptation can be used to solve the problem of data distribution mismatch. For example, knowledge learned from bulk-seq data can be transferred to scRNA-seq data or patient data to predict drug response in single cells or patients \cite{chen2022deep,lotfollahi2019scgen,ma2021few} which has important clinical significance.

\section{Cycle-consistent linear modeling}
\subsection{Encoder-decoder architecture}
\begin{figure*}[htb]
	\centering
	\includegraphics[scale=0.25]{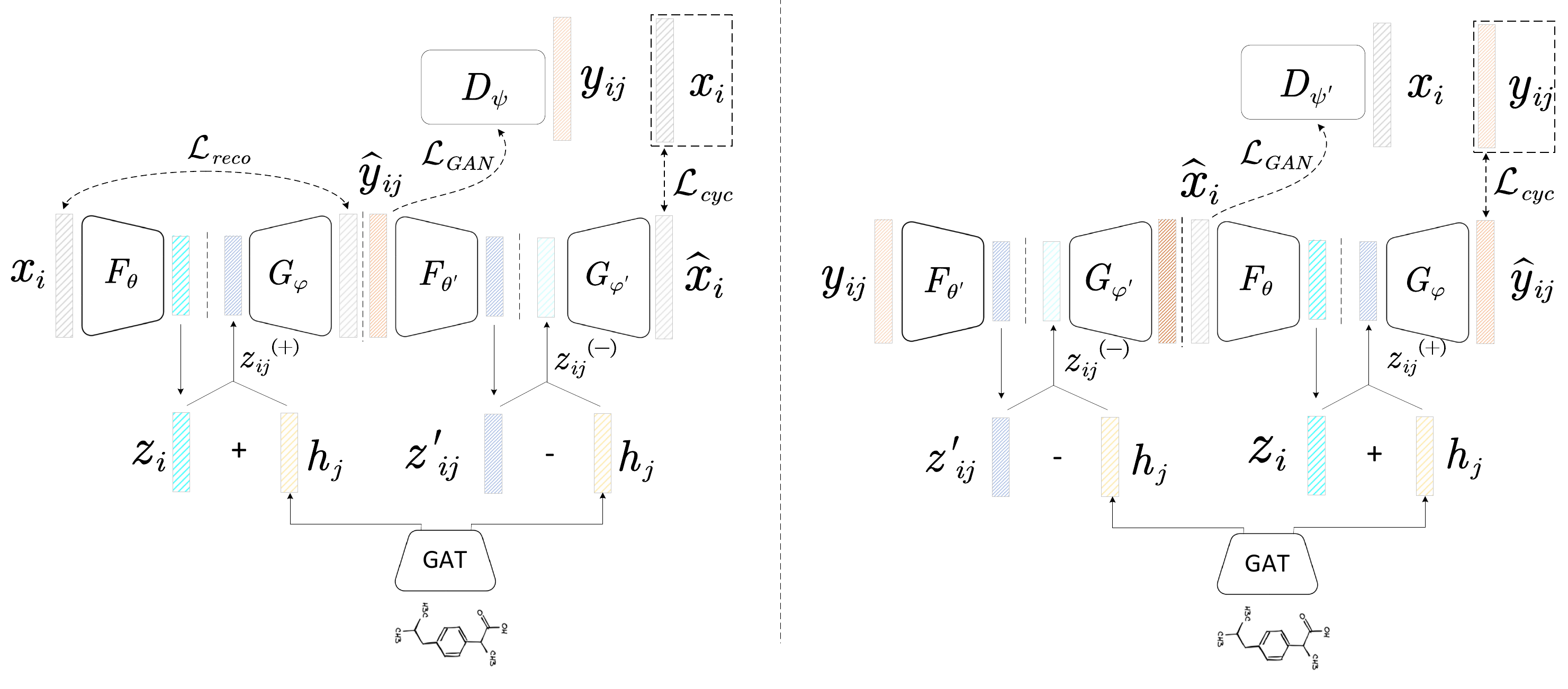}
	\caption{The illustrative diagram of our learning framework. Our model consists of two autoencoders, and they collaboratively learn and mutually improve each other’s predictive performance. The effect of drugs on cell state is linear model in the latent space so that our model is interpretable.}
	\label{fig:framwork}
\end{figure*}

We employ an encoder-decoder architecture to integrate the linear model and cycle-consistent constraints to predict cellular drug response. For simplicity, we refer to the proposed method as cycleCDR. Figure \ref{fig:framwork} shows the illustrate diagram of our proposed learning framework. The molecular signatures standing for cellular state serve as the input of an encoder, which maps the cellular features of control samples into a latent space. The decoder endeavors to yield expected cellular states, depending on the embedding to be decoded is unperturbed or perturbed by drugs in the latent space.

Formally, we define the encoder as $F_\theta$, which maps the cell state $x\in R^m$ to an $l$-dimensional latent vector $z\in R^l$. The decoder is defined as $G_\varphi$ that convert the latent vector into output space. Denote by $x$ and $y$ the initial (untreated cells) and perturbed cellular state (treated cells), and $z$ represents the mapped latent representation whose dimension is equal to the size of autoencoder bottleneck layer. Our encoder/decoder networks are fully or densely connected neural networks with rectified linear unit (ReLU) activation function, $\theta$ and $\varphi$ are the learnable parameters of the encoder and decoder. First, we require that the encoder-decoder architecture functions as an autoencoder to map the input molecular signatures to latent vector, and then recover the input signals. We optimize their parameters to minimize the reconstruction loss as follows:
\begin{equation}
    \mathcal{L}_{reco}=\sum_{i=1}^N ||x_i-G_\varphi(F_\theta(x_i))||_2^2
\end{equation}

The encoder converts the cellular states into low-dimensional but informative representations in the latent space. More importantly, in the latent space we can linearly model the cellular respose to drug perturbation (see Subsection \ref{sec:linear}).

\subsection{Drug perturbation encoder}
\label{sec:gat}

The graph attention network (GAT) was used to encode the drug perturbation into representation in the latent space. We used the SMILES of the drug to obtain the molecular graph $\mathcal{M}=<V,E>$, where $V$ is the set of nodes (atoms) and $E$ is the set of edges (chemical bonds). Assuming that $h_i$ is the embedding of node $i$ and $W$ is a learnable weight matrix, the attention score $\alpha_{ij}$ between node $i$ and its first-order neighbor node $j$ can be calculated using the following equation:
\begin{equation}
    \alpha_{ij}=\frac{\exp(elu(a^T(Wh_i,Wh_j)))}{\sum_{k\in \mathcal{N}(i)}\exp(elu(a^T(Wh_i,Wh_k)))}
\end{equation}
where $a$ is a learnable vector, $elu$ is the exponential linear unit activation function, and $\mathcal{N}(i)$represents the first-order neighbor of node $i$. The attention score $\alpha_{ij}$ was actually the softmax normalized message between node $i$ and its neighbors. Once the attention scores were computed, the output feature of node $i$ was computed by aggregating its neighbor features weighted by corresponding attention scores:
\begin{equation}
    h_i=\sigma(a_{ii}Wh_i+\sum\limits_{j\in N(i)}\alpha_{ij}Wh_j)
\end{equation}
where $\sigma(.)$ is the ReLU activation function.

\subsection{Linear modeling in latent space} \label{sec:linear}
Inspired by linear model in latent space, we assume that the effect of drug perturbation on cellular states follow the linear additivity in the latent space. Given that the cell state and drug disturbance are mapped to the same latent space, we construct a linear and easy-to-interpret model of the cellular drug response. Assuming that the $j$-th drug is mapped to $h_j$ via the drug perturbation encoder, the drug-induced cellular response in the latent space is defined as:
\begin{equation}
    {z}_{ij}^{(+)}={z}_i+{h}_j
\end{equation}
in which ${z}_{ij}^{(+)}$ represents the perturbed cellular state of the $i$-th cell by $j$-th drug in the latent space. Assuming that the actual cell state induced by the drug is $y_{ij}$, the decoder $G_\varphi$ should yield $\hat{y}_{ij}=G_\varphi({z}_i+{h}_j)$  that approximates to $y_{ij}$ as close as possible.

Based on the linear additivity, the cellular response to drug perturbation can be modeled from the opposite direction. By subtracting the drug representation  in the latent space, the drug effect should be removed and the perturbed cellular state could be restored to the perturbation-free state. Assume the encoder $F_{\theta^{'}}$ maps the actual cell response $y_{ij}$ to a vector $z_{ij}'=F_{\theta '}(y_{ij})$ into the common latent space, elimination of the drug interference can be directly subtracted by $h_j$. The restored representation of cellular state in the latent space can be defined as:
\begin{equation}
    {z}_{ij}^{(-)}={z}_{ij}'-{h}_j
\end{equation}
where $z_{ij}^{(-)}$ represents the latent representation of the $i$-th cell type (single cell) eliminating the perturbation of the $j$-th drug. Accordingly, another decoder $G_{\varphi '}$ is used to map $z_{ij}^{(-)}$ back to the cellular state that should be as close as possible to $x_i$.

For scenarios with paired data (unperturbed vs. perturbed), we use the mean squared error loss function as below:
\begin{equation}
\begin{split}
    \mathcal{L}_{{MSE}}=\frac{1}{N*K}\sum_{i=1}^N\sum_{j=1}^K [(y_{ij}-G_\varphi(z_i+h_j))^2\\
    +(x_i-G_{\varphi'}({z}_{ij}'-{h}_j))^2]
\end{split}
\end{equation}
The first term corresponds to loss in predicting the drug-perturbed cellular state from the initial state, while the second term corresponds to the loss in restoring the initial state from the drug-perturbed cell state.

However, for single-cell transcriptional response to drug perturbation, we lose the one-to-one correspondence of individual cells before and after drug treatment, because the cell body is destroyed in the single-cell RNA sequencing assay. Therefore, we leverage adversarial learning to align the data distribution between two domains (unperturbed state vs. perturbed state). In this situation, the decoder $G_\varphi$ functions as a generator that produces the perturbed cellular states. We introduce a discriminator $D_\psi$ to distinguish between actual and generated cellular responses. The adversarial loss is defined as follows:
\begin{equation}
\begin{split}
    \mathcal{L}_{{GAN}}(G_{\varphi},{D}_{\psi},X,Y)={E}_{\psi}[\log{D}_{\psi}(y_{ij})]\\+{E}_{\varphi,\theta}[\log(1-{D}_{\psi}(G_{\varphi}({z}_i+{h}_j)))]
\end{split}
\end{equation}
where the generator $G_{\varphi}$ tries to produce cellular response $G_{\varphi}({z}_i+{h}_j)$ that look similar to actual response, while $D_{\psi}$ aims to distinguish between generated samples and real samples. $G_{\varphi}$ aims to minimize this objective against an adversary $D_{\psi}$ that tries to maximize it, i.e., $\min_{G_\varphi}\max_{D_\psi}{L}_{GAN}(G_{\varphi},{D}_{\psi},X,Y)$. Correspondingly, another discriminator $D_{\psi '}$ is introduced to distinguish the actual initial state from generated cell states by $G_{\varphi '}$, and adversarial loss is defined as follows:
\begin{equation}
\begin{split}
    \mathcal{L}_{GAN}(G_{\varphi'},{D}_{\psi'},{Y},{X})={E}_{\psi'}[\log{D}_{\psi'}(x_i)]\\
    +{E}_{\varphi',\theta'}[\log(1-{D}_{\psi'}(G_{\psi'}({z}_{ij}'-{h}_j))]
\end{split}
\end{equation}

\subsection{Cycle-consistent loss}
Although adversarial learning-based domain adaption could align the data distribution of source domain and target domain, but a encoder network has adequate capacity to map a specific set of cellular states to any random permutation of cells within the target domain. Any of the learned mappings can induce an output distribution that matches the target distribution. Thus, adversarial losses alone cannot guarantee that the learned function can map an individual input $x_i$ to a desired output $y_i$. To further reduce the space of possible mappings, we require that the learned mapping functions should be cycle-consistent.

Formally, the initial cellular state $x_i$ is mapped to the a drug-perturbed state $\hat{y}_ij$ by drug $j$, which should be used to restore the initial state. We thus require $x_i\xrightarrow{{G}_{\varphi}(F_\theta(x_i)+h_j)}\widehat{y}_{ij}\xrightarrow{{G}_{\varphi '}(F_{\theta '}(\widehat{y}_{ij})-h_j)}\widehat{x}_i\approx x_i$.  Correspondingly, the perturbed cellular state is mapped to an unperturbed state that should be mapped to the corresponding perturbed state, namely, we require $y_{ij}\xrightarrow{G_{\varphi'}(F_{\theta'}(y_{ij})-h_j)}\widehat{x}_i\xrightarrow{G_{\varphi}(F_{\theta}(\hat{x_i})+h_j)}\widehat{y}_{ij}\approx y_{ij}$. Thus, we define the cycle-consistency loss function as follows:
\begin{equation}
    \mathcal{L}_{{cyc}}=\sum_{i=1}^N(x_i-\hat{x}_i)^2+\sum_{i=1}^N\sum_{j=1}^{K}(y_{ij}-\hat{y}_{ij})^2
\end{equation}

\subsection{Full objective}
For the transcriptional or proteomic responses measured on cell population, we have the paired data so that we define a relatively simple objective function as below:
\begin{equation}
    \mathcal{L}=\mathcal{L}_{reco}+\mathcal{L}_{MSE}
\end{equation}

For the single-cell cellular response to drug perturbation, we have no paired data and thus define the full objective function as:
\begin{equation}
    \mathcal{L}=\mathcal{L}_{reco}+\mathcal{L}_{GAN}+\lambda\mathcal{L}_{{cyc}}
\end{equation}
in which $\mathcal{L}_{GAN}=\mathcal{L}_{{GAN}}(G_{\varphi},{D}_{\psi},X,Y)+\mathcal{L}_{GAN}(G_{\varphi'},{D}_{\psi'},{Y},{X})$, $\lambda$ is the tradeoff parameter standing for the importance of cycle consistence constraints. In addition, similar to Taigman et al. [49], we regularize the generator produce nearly an identity mapping when real samples of the target domain (perturbed) are provided as the input to the generator. The identity loss is defined as $\mathcal{L}_{identity}=\sum_{i}^{N}\sum_{j}^{K}(G_{\varphi}(F_{\theta}(y))-y)^2|$ and implicitly used in the full objective.

The two autoencoders included in our model function collaboratively and mutually promote their predictive performance. In our ablation experiments, we attempted to use a single autoencoder for the cycle consistence leaning, namely $F_\theta=F_{\theta '}$ and $G_\varphi=G_{\varphi '}$, and found that their performance is comparative. We also evaluated the ablated model performance without reconstruction loss, as well as the ablated model only adversarial loss alone by removal of cycle consistency loss.

\section{Evaluation experiments}
\subsection{Experimental settings}
All the encoders and decoders for gene expression profiles were implemented using a multi-layer MLP with Relu activation function and batch normalization. The discriminator used in adversarial learning is implemented using a multi-layer MLP to distinguish the domain of the gene expression profile. The dimension of the bottleneck layer was set to 128. The drug molecular graph encoder comprised of two layers of GAT with an additional fully connected layer to adjust the embedding dimension. The multi-head attention mechanism was applied to the first layer and the number of heads was set to 10. The global max pooling over the node-level features is used to obtain the graph embedding. The parameters of the GAT encoder are initialized using the pretrained model based on attention-wise masked graph contrastive learning \cite{liu2022attention}. The hyperparameter $\lambda$ is set to 10 that used in cycleGAN \cite{zhu2017unpaired}.

To evaluate the performance of our model to predict cellular responses to drug perturbations, we use the coefficient of determination $r^2$ as an evaluation metric. For the evaluation of transcriptomic response prediction, we compute the $r^2$ score and explained variance (EV) over all genes to evaluate the performance in predicting transcriptome-wide response. Besides, since the expression levels of most genes in the perturbed cells remains similar to their control state, the performance metrics based on all genes are high but does not reflect
the true predictive capacity of a model. In contrast, the differentially expressed genes (DEGs) can more faithfully reflect the actual effect of drug interference on cellular state. Therefore, we also calculated the $r^2$ and EV metrics based on 50 most significantly differentially expressed genes. 

\subsection{Performance evaluation on bulk transcriptional response}
The L1000 dataset is a large collection of gene expression profiles that measure the responses of human cell lines to various compounds. It consists of approximately 1,400,000 gene expression profiles on the responses of about 50 human cell lines to one of about 20,000 compounds across a range of concentrations. We obtained data from the L1000 website, and used only the expression profiles treated by 10$\mu$M drug concentrations.  The gene expression values of technical replicates were averaged. As a result, we obtained drug response expression profiles spanning 17,775 drugs and 42 cell line types, including a total of 45,763 expression profiles over 978 landmark genes.

\begin{figure}[htbp]
    \centering
\includegraphics[width=0.5\textwidth]{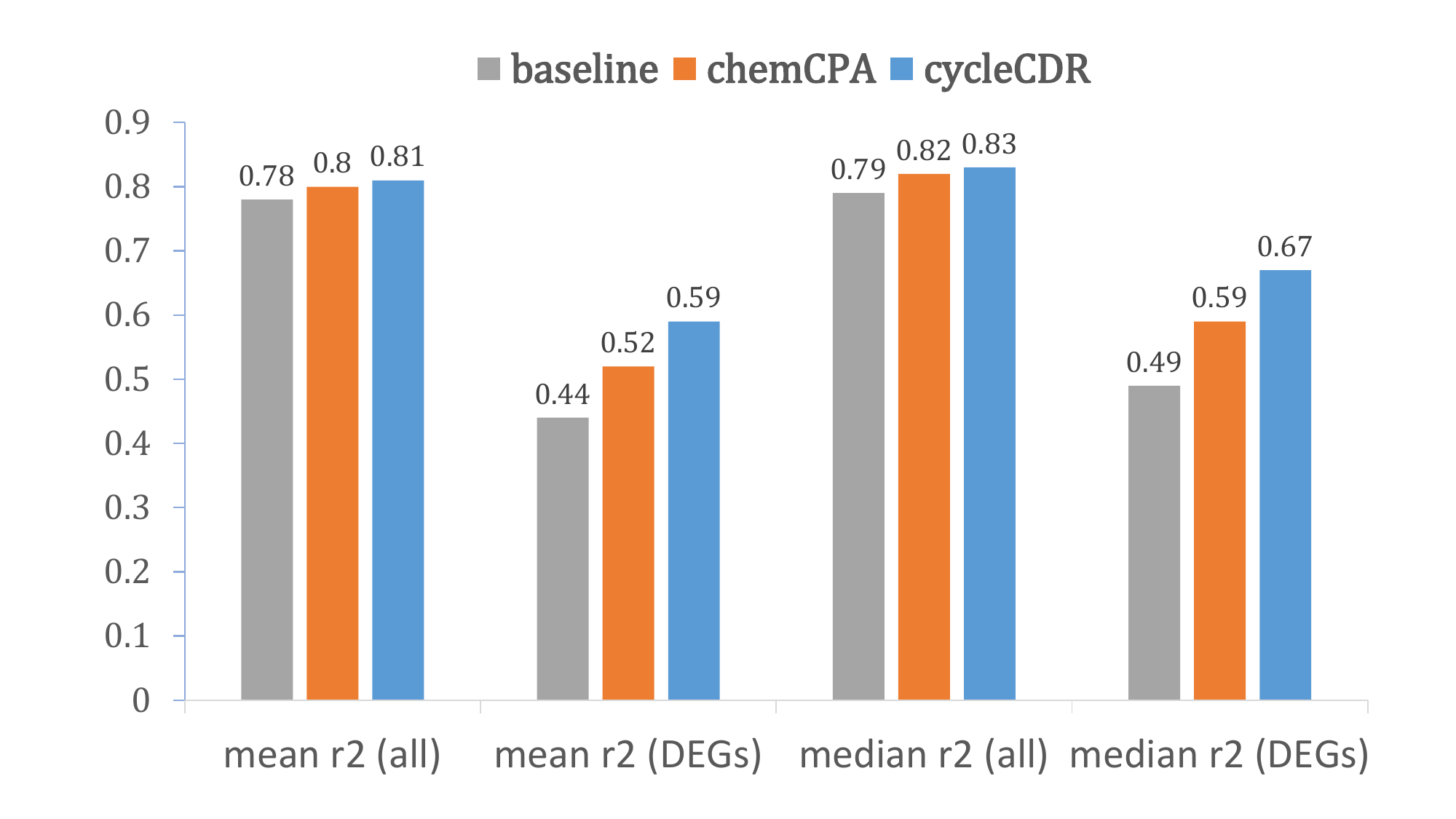}
\caption{Performance comparison of cycleCDR with the baseline model and chemCPA method on L1000 bulk transcriptional response dataset. }  \label{fig:rsqaure4L1000}
\end{figure}

\begin{figure*}[htbp]
 \begin{minipage}{1\textwidth}
    \centering
        \subfloat[]{\includegraphics[width=0.38\textwidth]
        {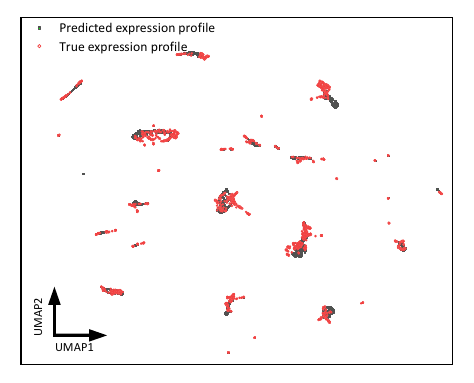}}
        \subfloat[]{\includegraphics[scale=0.5]{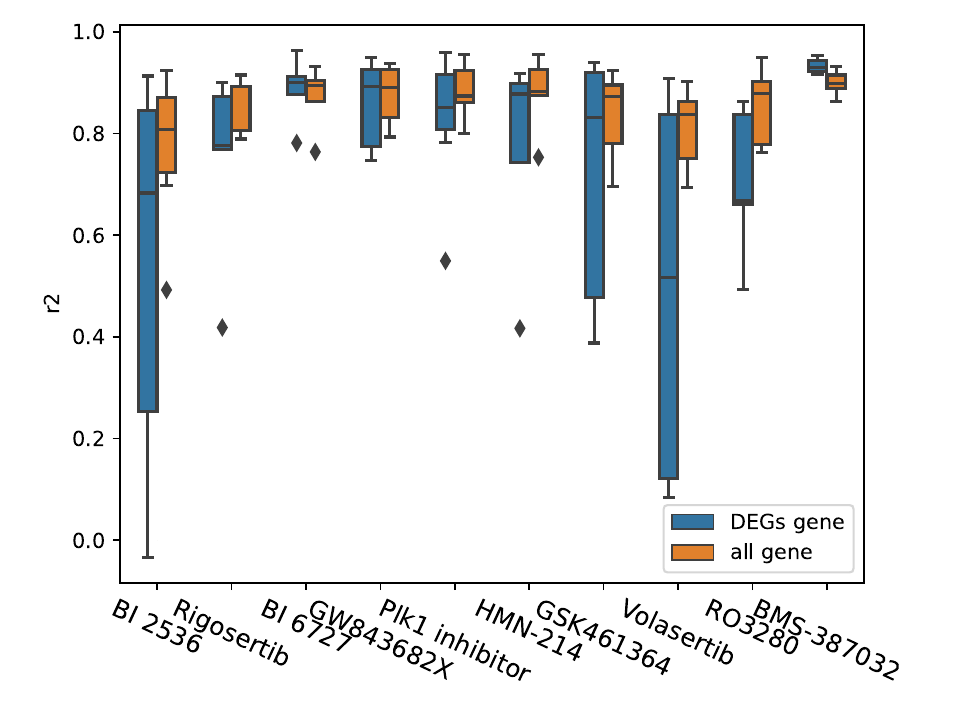}}
 \end{minipage}
\caption{Performance evaluation on L1000 bulk transctiptional response data. (a) UMAP visualization of predicted and actual expression profiles of 2,289 test samples. (b) The boxplots of $r^2$ scores regrading 10 drugs having the most number of perturbed cellular responses included in the test set.}  \label{fig:l1000}
\end{figure*}

\begin{figure}[htbp]
    \centering
\includegraphics[width=0.48\textwidth]{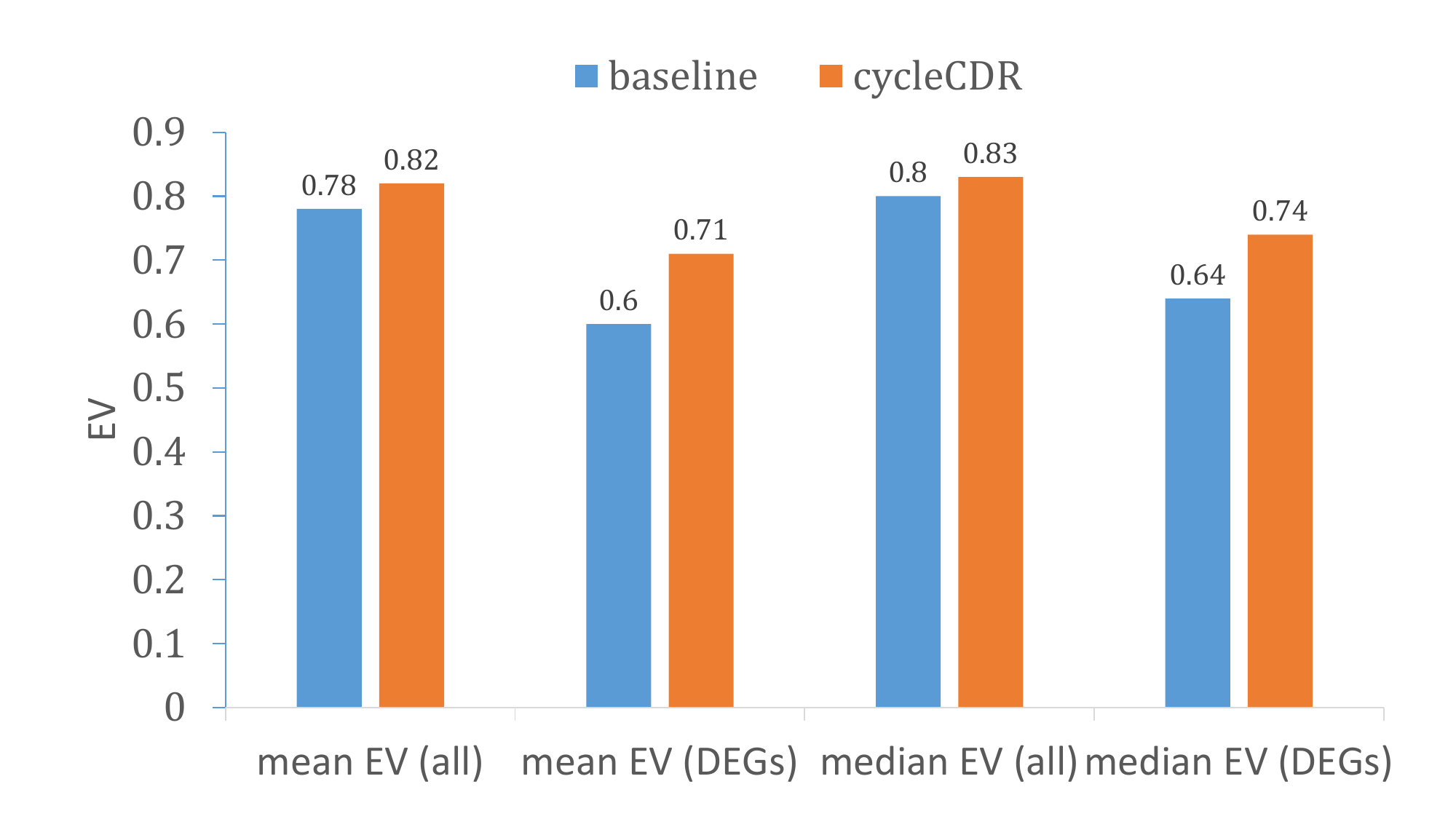}
\caption{Explained variance comparison of cycleCDR with baseline model on L1000 bulk transcriptional response dataset. }  \label{fig:EV4L1000}
\end{figure}

To benchmark our model performance, we constructed a baseline model that directly calculates the $r^2$ scores by discarding all perturbation information. This baseline has been also adopted by previous study \cite{hetzel2022predicting}. Also, we compared our model with another perturbation prediction model chemCPA. Figure \ref{fig:rsqaure4L1000} shows the mean and median $r^2$ scores obtained by each model on the test set. Compared to the baseline and chemCPA, our model achieved the highest $r^2$ scores on all genes and DEG set. The result strongly supports that our model achieves the current state-of-the-art level. To visually demonstrate the predictive ability of our model, we used the UMAP method to visualize the actual gene expression profiles and predicted gene expression profiles, as shown in Figure \ref{fig:l1000} (a). It can be seen that our model's predicted values are very close to the actual values, indicating that our model effectively captures the effect of drug perturbation on gene expression. Moreover, Figure \ref{fig:EV4L1000} shows the explained variance of the baseline and our model, Figure \ref{fig:l1000} (b) visualize the boxplots of $r^2$ scores regrading 10 drugs with the most perturbed cellular responses in the test set. The boxplots are grouped by drugs, and each group contains the prediction results of the model for different cell lines under the same drug perturbation.

\subsection{Performance evaluation on proteomic response}
Most molecular and targeted drugs achieve their pharmacological effects by affecting the function of proteins and protein complexes. The CPPA portal provides a set of large-scale proteomic expression levels measured by RPPA assays. The CPPA dataset includes 549 clinically relevant protein levels of 126 human cell lines perturbed by 99 drugs, which enable us to evaluate our model in predicting the proteomic response to drug perturbation.

\begin{figure}[htbp]
    \centering
\includegraphics[width=0.48\textwidth]{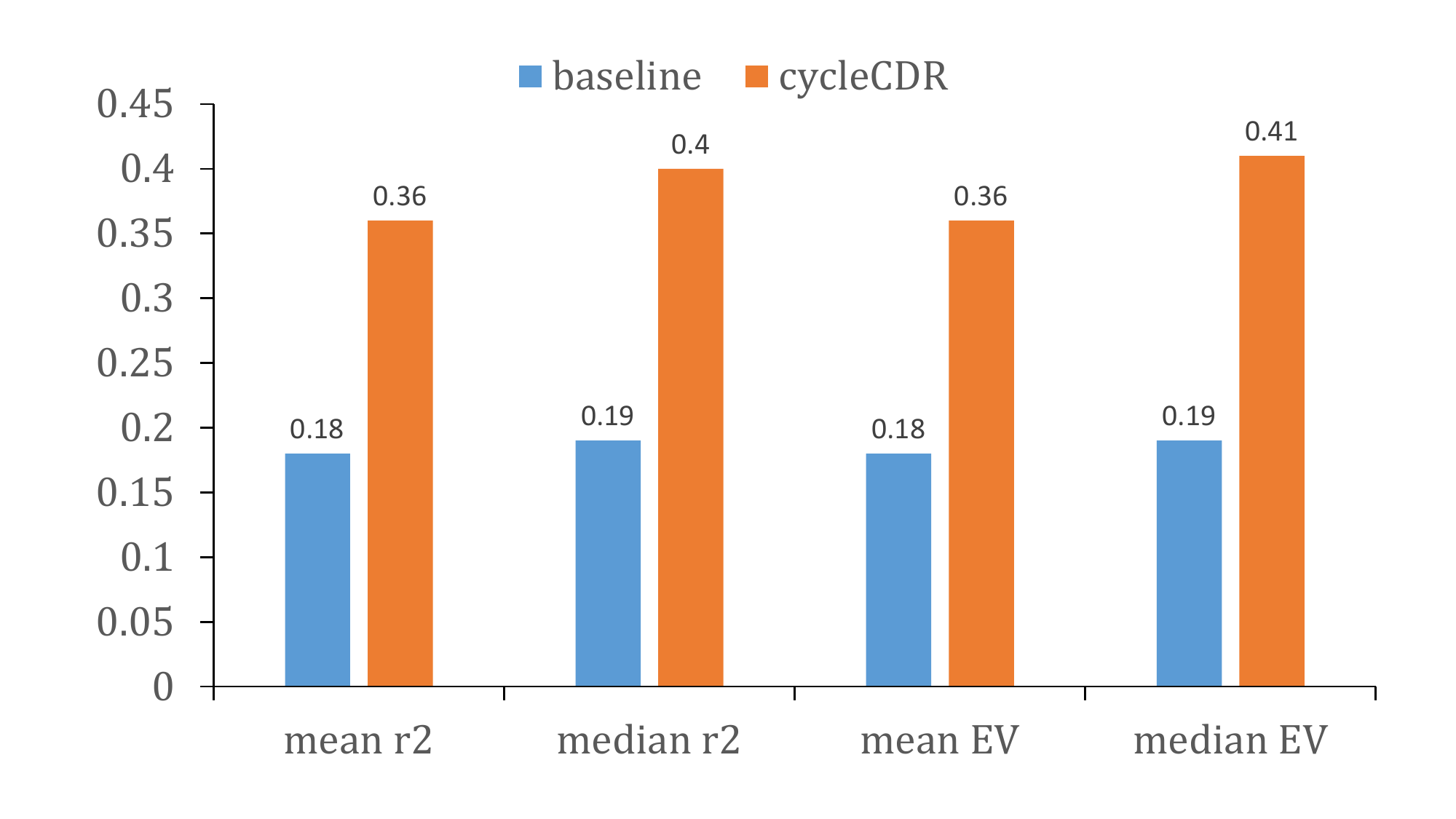}
\caption{Performance evaluation on bulk proteomic response dataset (CPPA).}  \label{fig:cppa-r2}
\end{figure}

We discard the data without dosage information and obtained 1,760 drug-cell line combinations spanning 538 proteins. The processed dataset were randomly divided into training ($n$=1,408), validation ($n$=246), and test ($n$=106) sets. As the CPPA data contains the measured proteomic profiles upon different drug dosage, we added an encoder for drug dosage to convert the drug dosage into an embedding and perform element-wise multiplication with the drug embedding. We show in Figure \ref{fig:cppa-r2}  the achieved mean and median $r^2$ and EV metrics on the test set. Compared to the baseline model, our model showed significant performance superiority. 

\subsection{Performance evaluation on single-cell transcriptional response}
\begin{figure}[htbp]
    \centering
\includegraphics[width=0.48\textwidth]{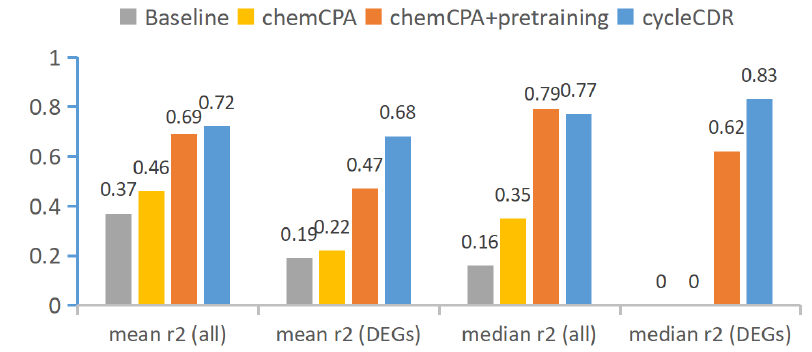}
\caption{Performance comparison of on sci-Plex3 single-cell transcriptional response dataset.}  \label{fig:r2_sciplex}
\end{figure}

\begin{figure*}[htbp]
 \begin{minipage}{1\textwidth}
    \centering
        \subfloat[]{\includegraphics[width=0.39\textwidth]{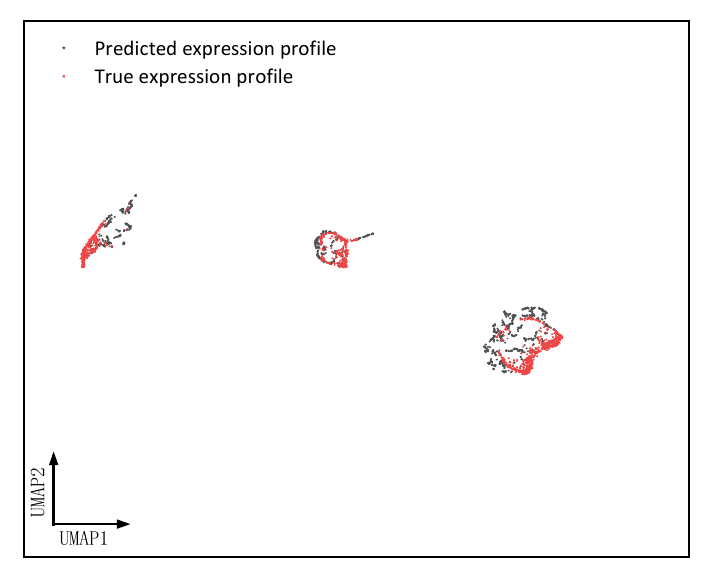}}
        \subfloat[]{\includegraphics[scale=0.5]{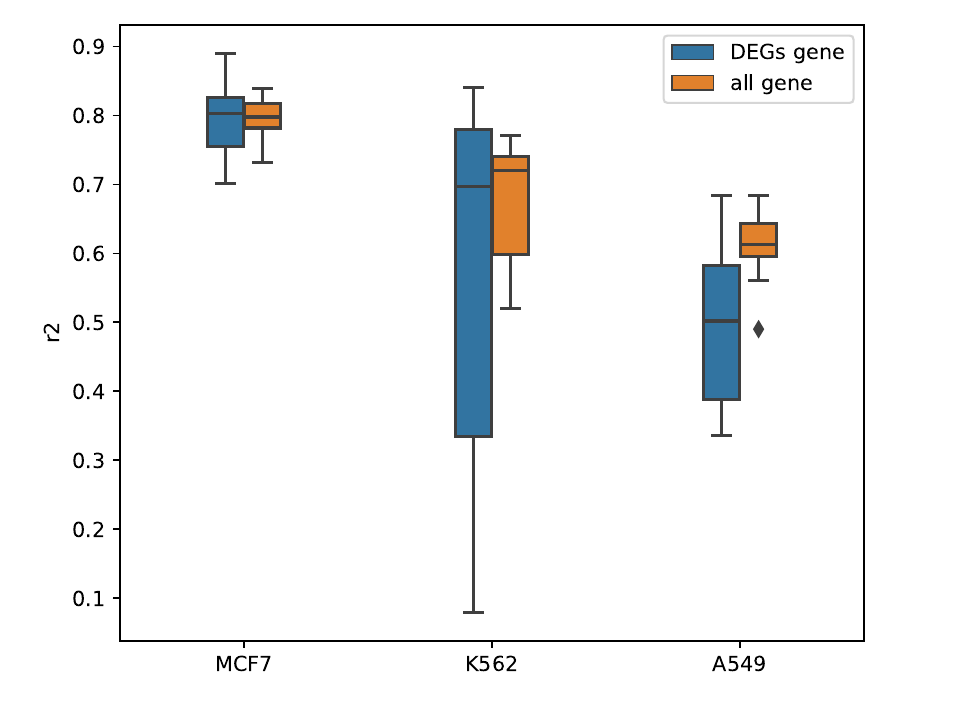}}
 \end{minipage}
\caption{Performance evaluation on sci-plex3 single-cell transctiptional response data. (a) UMAP visualization of predicted and actual expression profiles. (b) The boxplots of $r^2$ scores regrading three cell lines.} \label{fig:sci-plex3}
\end{figure*}

The sci-Plex3  single-cell resolution transcriptome response dataset has measured the transcriptional response of 3 human cancer cell lines to 188 compounds through high-throughput screening. For performance comparison, we used the sci-Plex3 dataset processed by chemCPA to evaluate the predictive ability of the model. Figure \ref{fig:r2_sciplex} shows the mean and median $r^2$ scores of our model on the dataset. Compared to both chemCPA with and without pretraining on L1000 bulk data, our model significantly achieved better performance. As shown in Figure \ref{fig:sci-plex3}, the UMAP visualization verified that the predicted transcriptional profiles are very close to the actual ones, indicating that our model can effectively capture single-cell transcriptional drug response. Figure \ref{fig:sci-plex3} shows the obtained $r^2$ scores of three major cell lines upon drug perturbations.

\subsection{Model ablation}
We first evaluate the effect of single autoencoder or dual autoencoder architecture on model performance. Table \ref{tab:ablation} shows the results of model ablation experiments performed on the sci-plex3 dataset. Although the dual autoencoder structure can improve the performance, its improvement is relatively weak. In addition, we also examined the effects of reconstruction loss and cycle consistence loss on the model separately, and found that cycle consistence loss contributes significantly to improve the model, especially for the prediction of DEGS. Most interestingly, when using the single autoencoder the combination usage of reconstruction loss and cycle consistence loss actually reduces the performance of the model, which may be caused by the limited expressive capacity of single autoencoder.


\begin{table*}[!ht]
    \centering
    \caption{Performance evaluation of ablated models} \label{tab:ablation}
    \begin{tabular}{|c|c|c|c|c|c|}
    \hline
        Single AE & Dual AE & Reco Loss & Cycle Loss & mean $r^2$ (all) & mean $r^2$ (DEGs) \\ \hline

        $\checkmark$ & ~ & $\checkmark$ & ~ & 0.53 & 0.61 \\ \hline
        $\checkmark$ & ~ & ~ & $\checkmark$ & 0.54 & 0.63 \\ \hline
        $\checkmark$ & ~ & $\checkmark$ & $\checkmark$ & 0.58 & 0.67 \\ \hline
        ~ & $\checkmark$ & $\checkmark$ & ~ & 0.65 & 0.59 \\ \hline
        ~ & $\checkmark$  & ~ & $\checkmark$ & 0.66 & 0.66 \\ \hline
        ~ & $\checkmark $& $\checkmark$ & $\checkmark$ & 0.72 & 0.68 \\ \hline
    \end{tabular}
\end{table*}

\section{Conclusion}
In this paper, we proposed a novel deep network model to learn cellular response to drug perturbations. We integrated the cycle consistency loss and linear model in latent space into a end-to-end learning framework, which enables our model to learn expressive and transferable drug representations. We evaluate the proposed model on both bulk and single-cell transcriptional responses, as well as a proteomic drug response dataset. The experimental results on these benchmark datasets demonstrate the superior performance of our model.

\bibliographystyle{unsrt}
\bibliography{reference}

\begin{thebibliography}{31}
\providecommand{\natexlab}[1]{#1}

\bibitem[{Aissa et~al.(2021)Aissa, Islam, Ariss, Go, Rader, Conrardy, Gajda,
  Rubio-Perez, Valyi-Nagy, Pasquinelli et~al.}]{aissa2021single}
Aissa, A.~F.; Islam, A.~B.; Ariss, M.~M.; Go, C.~C.; Rader, A.~E.; Conrardy,
  R.~D.; Gajda, A.~M.; Rubio-Perez, C.; Valyi-Nagy, K.; Pasquinelli, M.; et~al.
  2021.
\newblock Single-cell transcriptional changes associated with drug tolerance
  and response to combination therapies in cancer.
\newblock \emph{Nature communications}, 12(1): 1628.

\bibitem[{Arjovsky, Chintala, and Bottou(2017)}]{arjovsky2017wasserstein}
Arjovsky, M.; Chintala, S.; and Bottou, L. 2017.
\newblock Wasserstein generative adversarial networks.
\newblock In \emph{International conference on machine learning}, 214--223.
  PMLR.

\bibitem[{Chawla et~al.(2022)Chawla, Rockstroh, Lehman, Ratther, Jain, Anand,
  Gupta, Bhattacharya, Poonia, Rai et~al.}]{chawla2022gene}
Chawla, S.; Rockstroh, A.; Lehman, M.; Ratther, E.; Jain, A.; Anand, A.; Gupta,
  A.; Bhattacharya, N.; Poonia, S.; Rai, P.; et~al. 2022.
\newblock Gene expression based inference of cancer drug sensitivity.
\newblock \emph{Nature communications}, 13(1): 5680.

\bibitem[{Chen et~al.(2019)Chen, Chen, Jiang, and Jin}]{chen2019joint}
Chen, C.; Chen, Z.; Jiang, B.; and Jin, X. 2019.
\newblock Joint domain alignment and discriminative feature learning for
  unsupervised deep domain adaptation.
\newblock In \emph{Proceedings of the AAAI conference on artificial
  intelligence}, volume~33, 3296--3303.

\bibitem[{Chen et~al.(2022)Chen, Wang, Ma, Wang, Liu, Li, Xu, and
  Ma}]{chen2022deep}
Chen, J.; Wang, X.; Ma, A.; Wang, Q.-E.; Liu, B.; Li, L.; Xu, D.; and Ma, Q.
  2022.
\newblock Deep transfer learning of cancer drug responses by integrating bulk
  and single-cell RNA-seq data.
\newblock \emph{Nature Communications}, 13(1): 6494.

\bibitem[{Gao, Yang, and Wang(2021)}]{gao2021collaborative}
Gao, L.-G.; Yang, M.-Y.; and Wang, J.-X. 2021.
\newblock Collaborative matrix factorization with soft regularization for
  drug-target interaction prediction.
\newblock \emph{Journal of Computer Science and Technology}, 36: 310--322.

\bibitem[{He et~al.(2022)He, Liu, Wu, and Xie}]{he2022context}
He, D.; Liu, Q.; Wu, Y.; and Xie, L. 2022.
\newblock A context-aware deconfounding autoencoder for robust prediction of
  personalized clinical drug response from cell-line compound screening.
\newblock \emph{Nature Machine Intelligence}, 4(10): 879--892.

\bibitem[{Hetzel et~al.(2022)Hetzel, Boehm, Kilbertus, G{\"u}nnemann, Theis
  et~al.}]{hetzel2022predicting}
Hetzel, L.; Boehm, S.; Kilbertus, N.; G{\"u}nnemann, S.; Theis, F.; et~al.
  2022.
\newblock Predicting cellular responses to novel drug perturbations at a
  single-cell resolution.
\newblock \emph{Advances in Neural Information Processing Systems}, 35:
  26711--26722.

\bibitem[{Hetzel et~al.(2021)Hetzel, Fischer, G{\"u}nnemann, and
  Theis}]{hetzel2021graph}
Hetzel, L.; Fischer, D.~S.; G{\"u}nnemann, S.; and Theis, F.~J. 2021.
\newblock Graph representation learning for single-cell biology.
\newblock \emph{Current Opinion in Systems Biology}, 28: 100347.

\bibitem[{Jia et~al.(2021)Jia, Hu, Pei, Dai, Wang, and Zhao}]{jia2021deep}
Jia, P.; Hu, R.; Pei, G.; Dai, Y.; Wang, Y.-Y.; and Zhao, Z. 2021.
\newblock Deep generative neural network for accurate drug response imputation.
\newblock \emph{Nature Communications}, 12(1): 1740.

\bibitem[{Liu et~al.(2022)Liu, Huang, Liu, and Deng}]{liu2022attention}
Liu, H.; Huang, Y.; Liu, X.; and Deng, L. 2022.
\newblock Attention-wise masked graph contrastive learning for predicting
  molecular property.
\newblock \emph{Briefings in Bioinformatics}, 23(5): bbac303.

\bibitem[{Long et~al.(2018)Long, Cao, Wang, and Jordan}]{long2018conditional}
Long, M.; Cao, Z.; Wang, J.; and Jordan, M.~I. 2018.
\newblock Conditional adversarial domain adaptation.
\newblock \emph{Advances in neural information processing systems}, 31.

\bibitem[{Lopez, Gayoso, and Yosef(2020)}]{lopez2020enhancing}
Lopez, R.; Gayoso, A.; and Yosef, N. 2020.
\newblock Enhancing scientific discoveries in molecular biology with deep
  generative models.
\newblock \emph{Molecular Systems Biology}, 16(9): e9198.

\bibitem[{Lotfollahi et~al.(2021)Lotfollahi, Susmelj, De~Donno, Ji, Ibarra,
  Wolf, Yakubova, Theis, and Lopez-Paz}]{lotfollahi2021learning}
Lotfollahi, M.; Susmelj, A.~K.; De~Donno, C.; Ji, Y.; Ibarra, I.~L.; Wolf,
  F.~A.; Yakubova, N.; Theis, F.~J.; and Lopez-Paz, D. 2021.
\newblock Learning interpretable cellular responses to complex perturbations in
  high-throughput screens.
\newblock \emph{BioRxiv}, 2021--04.

\bibitem[{Lotfollahi, Wolf, and Theis(2019)}]{lotfollahi2019scgen}
Lotfollahi, M.; Wolf, F.~A.; and Theis, F.~J. 2019.
\newblock scGen predicts single-cell perturbation responses.
\newblock \emph{Nature methods}, 16(8): 715--721.

\bibitem[{Ma et~al.(2021)Ma, Fong, Luo, Bakkenist, Shen, Mourragui, Wessels,
  Hafner, Sharan, Peng et~al.}]{ma2021few}
Ma, J.; Fong, S.~H.; Luo, Y.; Bakkenist, C.~J.; Shen, J.~P.; Mourragui, S.;
  Wessels, L.~F.; Hafner, M.; Sharan, R.; Peng, J.; et~al. 2021.
\newblock Few-shot learning creates predictive models of drug response that
  translate from high-throughput screens to individual patients.
\newblock \emph{Nature Cancer}, 2(2): 233--244.

\bibitem[{Madhukar et~al.(2019)Madhukar, Khade, Huang, Gayvert, Galletti,
  Stogniew, Allen, Giannakakou, and Elemento}]{madhukar2019bayesian}
Madhukar, N.~S.; Khade, P.~K.; Huang, L.; Gayvert, K.; Galletti, G.; Stogniew,
  M.; Allen, J.~E.; Giannakakou, P.; and Elemento, O. 2019.
\newblock A Bayesian machine learning approach for drug target identification
  using diverse data types.
\newblock \emph{Nature communications}, 10(1): 5221.

\bibitem[{Pei et~al.(2018)Pei, Cao, Long, and Wang}]{pei2018multi}
Pei, Z.; Cao, Z.; Long, M.; and Wang, J. 2018.
\newblock Multi-adversarial domain adaptation.
\newblock In \emph{Thirty-second AAAI conference on artificial intelligence}.

\bibitem[{Robichaux et~al.(2021)Robichaux, Le, Vijayan, Hicks, Heeke, Elamin,
  Lin, Udagawa, Skoulidis, Tran et~al.}]{robichaux2021structure}
Robichaux, J.~P.; Le, X.; Vijayan, R.; Hicks, J.~K.; Heeke, S.; Elamin, Y.~Y.;
  Lin, H.~Y.; Udagawa, H.; Skoulidis, F.; Tran, H.; et~al. 2021.
\newblock Structure-based classification predicts drug response in EGFR-mutant
  NSCLC.
\newblock \emph{Nature}, 597(7878): 732--737.

\bibitem[{Roohani, Huang, and Leskovec(2022)}]{roohani2022gears}
Roohani, Y.; Huang, K.; and Leskovec, J. 2022.
\newblock GEARS: Predicting transcriptional outcomes of novel multi-gene
  perturbations.
\newblock \emph{BioRxiv}, 2022--07.

\bibitem[{Sankaranarayanan et~al.(2018)Sankaranarayanan, Balaji, Castillo, and
  Chellappa}]{sankaranarayanan2018generate}
Sankaranarayanan, S.; Balaji, Y.; Castillo, C.~D.; and Chellappa, R. 2018.
\newblock Generate to adapt: Aligning domains using generative adversarial
  networks.
\newblock In \emph{Proceedings of the IEEE conference on computer vision and
  pattern recognition}, 8503--8512.

\bibitem[{Sharifi-Noghabi et~al.(2021)Sharifi-Noghabi, Harjandi, Zolotareva,
  Collins, and Ester}]{sharifi2021out}
Sharifi-Noghabi, H.; Harjandi, P.~A.; Zolotareva, O.; Collins, C.~C.; and
  Ester, M. 2021.
\newblock Out-of-distribution generalization from labelled and unlabelled gene
  expression data for drug response prediction.
\newblock \emph{Nature Machine Intelligence}, 3(11): 962--972.

\bibitem[{Srivatsan et~al.(2020)Srivatsan, McFaline-Figueroa, Ramani, Saunders,
  Cao, Packer, Pliner, Jackson, Daza, Christiansen
  et~al.}]{srivatsan2020massively}
Srivatsan, S.~R.; McFaline-Figueroa, J.~L.; Ramani, V.; Saunders, L.; Cao, J.;
  Packer, J.; Pliner, H.~A.; Jackson, D.~L.; Daza, R.~M.; Christiansen, L.;
  et~al. 2020.
\newblock Massively multiplex chemical transcriptomics at single-cell
  resolution.
\newblock \emph{Science}, 367(6473): 45--51.

\bibitem[{Subramanian et~al.(2017)Subramanian, Narayan, Corsello, Peck, Natoli,
  Lu, Gould, Davis, Tubelli, Asiedu et~al.}]{subramanian2017next}
Subramanian, A.; Narayan, R.; Corsello, S.~M.; Peck, D.~D.; Natoli, T.~E.; Lu,
  X.; Gould, J.; Davis, J.~F.; Tubelli, A.~A.; Asiedu, J.~K.; et~al. 2017.
\newblock A next generation connectivity map: L1000 platform and the first
  1,000,000 profiles.
\newblock \emph{Cell}, 171(6): 1437--1452.

\bibitem[{Tzeng et~al.(2019)Tzeng, Hoffman, Zhang, Saenko, and
  Darrell}]{tzeng2019deep}
Tzeng, E.; Hoffman, J.; Zhang, N.; Saenko, K.; and Darrell, T. 2019.
\newblock Deep domain confusion: Maximizing for domain invariance. arXiv 2014.
\newblock \emph{arXiv preprint arXiv:1412.3474}.

\bibitem[{Wang, He, and Katabi(2020)}]{wang2020continuously}
Wang, H.; He, H.; and Katabi, D. 2020.
\newblock Continuously indexed domain adaptation.
\newblock \emph{arXiv preprint arXiv:2007.01807}.

\bibitem[{Yu and Welch(2022)}]{yu2022perturbnet}
Yu, H.; and Welch, J.~D. 2022.
\newblock PerturbNet predicts single-cell responses to unseen chemical and
  genetic perturbations.
\newblock \emph{bioRxiv}, 2022--07.

\bibitem[{Zhang, Lu, and Zang(2022)}]{zhang2022cnn}
Zhang, C.; Lu, Y.; and Zang, T. 2022.
\newblock CNN-DDI: a learning-based method for predicting drug--drug
  interactions using convolution neural networks.
\newblock \emph{BMC bioinformatics}, 23(1): 1--12.

\bibitem[{Zhao et~al.(2020)Zhao, Li, Chen, Luo, Ju, Nesser, Johnson-Camacho,
  Boniface, Lawrence, Pande et~al.}]{zhao2020large}
Zhao, W.; Li, J.; Chen, M.-J.~M.; Luo, Y.; Ju, Z.; Nesser, N.~K.;
  Johnson-Camacho, K.; Boniface, C.~T.; Lawrence, Y.; Pande, N.~T.; et~al.
  2020.
\newblock Large-scale characterization of drug responses of clinically relevant
  proteins in cancer cell lines.
\newblock \emph{Cancer cell}, 38(6): 829--843.

\bibitem[{Zheng et~al.(2017)Zheng, Fu, Mei, and Luo}]{zheng2017learning}
Zheng, H.; Fu, J.; Mei, T.; and Luo, J. 2017.
\newblock Learning multi-attention convolutional neural network for
  fine-grained image recognition.
\newblock In \emph{Proceedings of the IEEE international conference on computer
  vision}, 5209--5217.

\bibitem[{Zhu et~al.(2017)Zhu, Park, Isola, and Efros}]{zhu2017unpaired}
Zhu, J.-Y.; Park, T.; Isola, P.; and Efros, A.~A. 2017.
\newblock Unpaired image-to-image translation using cycle-consistent
  adversarial networks.
\newblock In \emph{Proceedings of the IEEE international conference on computer
  vision}, 2223--2232.

\end{thebibliography}

\end{document}